# *AirPen*: A Touchless Fingertip Based Gestural Interface for Smartphones and Head-Mounted Devices

**Varun Jain**
varun.in@tcs.com
TCS Research
New Delhi, India

**Ramya Hebbalaguppe**
ramya.hebbalaguppe@tcs.com
TCS Research
New Delhi, India

## ABSTRACT

Hand gestures are an intuitive, socially acceptable, and a non-intrusive interaction modality in Mixed Reality (MR) and smartphone based applications. Unlike speech interfaces, they tend to perform well even in shared and public spaces. Hand gestures can also be used to interact with smartphones in situations where the user's ability to physically touch the device is impaired. However, accurate gesture recognition can be achieved through state-of-the-art deep learning models or with the use of expensive sensors. Despite the robustness of these deep learning models, they are computationally heavy and memory hungry, and obtaining real-time performance on-device without additional hardware is still a challenge. To address this, we propose *AirPen*: an analogue to pen on paper, but in air, for in-air writing and gestural commands that works seamlessly in First and Second Person View. The models are trained on a GPU machine and ported on an Android smartphone. *AirPen* comprises of three deep learning models that work in tandem: MobileNetV2 for hand localisation, our custom fingertip regression architecture followed by a Bi-LSTM model for gesture classification. The overall framework works in real-time on mobile devices and achieves a classification accuracy of 80% with an average latency of only 0.12 *s*.

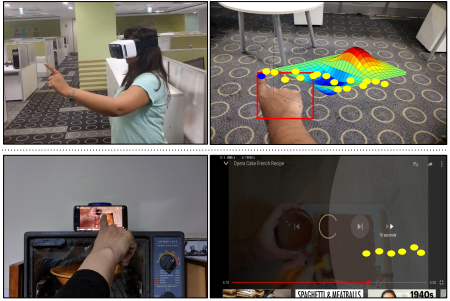

**Figure 1: Our framework works on commercial off-the-shelf smartphones and generic video see-through Head-Mounted Devices like the Google Cardboard.**

*(top, First Person View)* **Illustration of our MR application on Google Cardboard for Android devices: The application shows enhanced visualisation of data in the form of 3D graphs. Manipulation of the graph is done through pointing hand gestures like a left swipe as visualised by the gesture trail (in yellow).**

*(bottom, Second Person View)* **Pointing hand gesture to interact with the Android YouTube application: a user whose hands are soiled in flour while baking a cake tries to skip forward a YouTube instructional video on baking by a simple right swipe gesture.**

## KEYWORDS

intelligent user interfaces, activity/ hand gesture recognition, edge computing, smartphones

## INTRODUCTION

Over the past few decades, technology has transitioned from desktop to mobile computing. With this transition, the way we interact with devices also has evolved from keyboard/mice to touch and speech interfaces. However, despite of being a seemingly natural means of interaction with the virtual world, the use of hands as an input modality is still not mainstream. Over the past years, depth sensing solutions such as Leap Motion have become the de facto approach in solving the problem whereas monocular marker-less hand tracking has received relatively less attention. While such a solution could help in mass-market adoption since it does not require additional hardware, doing so with efficient real-time performing models that can accurately localise hands and fingertips is challenging.

Hand gestures can be effective in reducing the effects of many Situationally-Induced Impairments and Disabilities (SIID) in Second Person View applications (Figure 1) [15]. Public displays that use touchscreens such as ATM machines, kiosks and in malls face issues with hygiene & cleaning requirements, robustness against extended use & potential damage, responsiveness, and finally reachability that compromise viewability and location. The problem of contamination is particularly critical in settings such as assembly and food processing plants, and in bio-tech research laboratories. Further, speech interfaces tend to be intrusive and less accurate in noisy outdoor environments to record experimental data and results. For example, during an industrial inspection and repair operation that involves multiple inspectors situated in close vicinity, machine noise and cross-talk can hamper speech recognition. In these situations, hand gestures can provide an effective solution which is intuitive while being socially acceptable.

In First Person View applications (Figure 1), expensive AR/MR devices such as the Microsoft HoloLens, Daqri and Meta Glasses provide a rich user interface by using recent hardware advancements. They are equipped with a variety of on-board sensors including multiple cameras, a depth sensor and proprietary processors. This makes them expensive and unaffordable for mass adoption.

We propose a computationally effective pointing hand gesture recognition framework that works without depth information and the need of specialised hardware, thereby providing mass accessibility of gestural interfaces to commodity smartphone devices and affordable video see-through HMDs. Since it can run on a regular System on Chip (SoC), it has the benefit of being able to work in remote environments without the need of internet connectivity or a remote GPU machine. Our framework works seamlessly in First Person View such as in the case of HMDs and in Second Person View applications such as interacting with displays that are mounted on a camera. Public kiosks, bio-tech laboratories, smartphones, industrial inspection and repair, tele-presence, and data visualisation are some of the immediate applications for our framework.

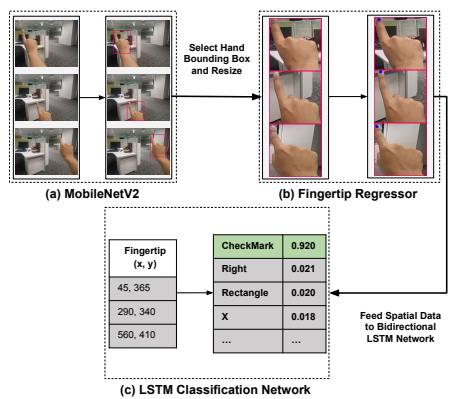

**Figure 2: *AirPen* pointing gestural framework consists of: *(a)* The *MobileNetV2* detector block that takes RGB image as an input and outputs a hand candidate bounding box, *(b)* we crop and resize the hand candidate and feed it to our Fingertip Regressor architecture (Figure 3) for localising the fingertip, and, *(c)* a Bi-LSTM network for the classification of fingertip detections on subsequent frames into different gestures.**

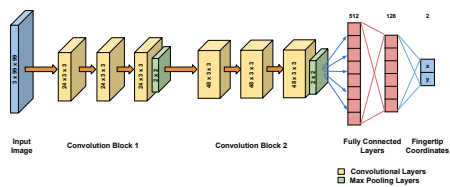

**Figure 3: Overview of our proposed fingertip regressor architecture for fingertip localisation. The input to our network are 3× 99×99 sized RGB images. Each of the 2 convolutional blocks have 3 convolutional layers each followed by a max-pooling layer. The 3 fully connected layers regress over fingertip spatial location.**

To demonstrate the generic nature of our framework, we evaluate the detection of 10 complex gestures performed using the pointing hand pose with a sample Android application. (Figure 1)

## RELATED WORK

The efficacy of hand gestures as an interaction modality for MR applications on smartphones and HMDs has been extensively explored in the past [7]. However, most work done towards bringing intuitive hand gestures to mobile applications was based on fingertip markers as in the case of SixthSense [10]. Such markers are often not user-friendly and might constitute to a SIID in themselves. Others state-of-the-art solutions are computationally expensive and rely on remote networked GPU machines for processing [4]. Most of the recent work focuses on depth data to recognise gestures in real time [16]. This limits the use-cases of the solution since it depends on specialised hardware such as a stereo-camera setup or an infrared projector. Hence, we focus on monocular RGB camera data to ensure mass adoption.

Accurate hand segmentation is very important in all gesture recognition applications. Li et al. [8] observed the response of *Gabor* filters to examine local appearance features in skin colour regions. Most early approaches are faced by the following challenges: *(i)* movement of the hand relative to the HMD renders the hand blurry, which makes it difficult to detect and track it, thereby impeding classification accuracy. *(ii)* sudden changes in illumination conditions and the presence of skin-like colours and texture in the background causes algorithms with skin feature dependency to fail. To this end, we look at utilising the current state-of-the-art object detection architectures like MobileNetV2 [14], YOLOv2 [12] and Faster R-CNN [13] for hand detection.

Several works [17] use depth information from sensors such as the Microsoft Kinect that restricts its applications in head mounted devices. Moreover, most depth sensors perform poorly in the presence of strong specular reflection, direct sunlight, incandescent light and outdoor environments due to the presence of infrared radiation [3].

## PROPOSED FRAMEWORK

There has been an increased emphasis on developing *end-to-end* networks that learn to model a number of sub-problems implicitly while training. While this has several advantages in learning joint tasks like object-detection followed by classification, it usually relies on the presence of a large amount of labelled data which has discriminative features useful for each of the sub-problems.

Hence, we propose an ensemble of architectures capable of recognising a variety of pointing hand gestures for frugal AR wearable devices with a monocular RGB camera input that requires only a limited amount of labelled classification data. The proposed architecture is capable of classifying fingertip motion patterns into different hand gestures. Figure 2 shows the building blocks of *AirPen*: *(i) MobileNetV2* [14] takes a single RGB image as an input and outputs a hand candidate bounding box. The input images are first down-scaled to $640 \times 480$ resolution to reduce processing time without compromising on the quality of image features. *(ii)* The detected hand candidates are then fed to a *fingertip regressor* as depicted in Figure 3 which outputs the spatial location of the fingertip. *(iii)* The collection of these is then fed it to the *Bi-LSTM* network for classifying the motion pattern into different gestures.

## EXPERIMENTS AND RESULTS

Since the framework comprises of three networks, we evaluate the performance of each of the networks individually to arrive at the best combination of networks for our proposed application. All results are reported on a Snapdragon® 845 SoC powered OnePlus 6.

For all the experiments pertaining to hand detection and fingertip localisation, we use the hand dataset [5]. Out of the 24 subjects present in the dataset, we choose 17 subjects' data for training with a validation split of 70:30, and 7 subjects' data (24, 155 images) for testing the networks.

## Hand Detection

Table 1 reports percentage of mean Absolute Precision (mAP) and frame rate for hand candidate detection. Even though MobileNetV2 [14] achieved higher frame-rate compared to others, it produced high false positives hence resulted in poor classification performance. It is observed that YOLOv2 can also run on-device although it outputs fewer frames as compared to MobileNetV2. Also, YOLOv2 performs poorly in localising the hand candidate at higher IoU that is required for including the fingertip. It is worth noticing that the model size for MobileNetV2 is significantly less than the rest of the models. It enables us to port the model on mobile device and removes the framework's dependence on a remote server. This helps reduce latency introduced by the network and can enable a enable a wider reach of frugal devices for MR applications.

## Fingertip Localisation

The $(x, y)$ finger coordinate estimated at the last layer are compared against ground truth values to compute rate of success with changing thresholds on the error (in pixels). We benchmark our model against the network proposed by Huang et al. [6]. The expected error for our model is lower and it achieves 89% accuracy with an error tolerance of 10 pixels on an input image of $99 \times 99$ resolution as compared to 83% for the benchmark.

| Model | On Device | mAP IoU=0.5 | mAP IoU=0.7 | rate (FPS) | Model Size |
|---|---|---|---|---|---|
| F-RCNN VGG16 | ✗ | 98.1 | 86.9 | 3 | 546 MB |
| F-RCNN VGG1024 | ✗ | 96.8 | 86.7 | 10 | 350 MB |
| F-RCNN ZF | ✗ | 97.3 | 89.2 | 12 | 236 MB |
| YOLOv2 | ✓ | 93.9 | 78.2 | 2 | 202 MB |
| **MobileNetV2** | ✓ | **89.1** | **85.3** | 9 | **12 MB** |

**Table 1: Performance of various methods on the *SCUT-Ego-Finger* dataset for hand detection. mAP score, frame-rate and the model size are reported with the variation in IoU.**

| Method | Precision | Recall | $F_1$ Score |
|--------|-----------|--------|-------------|
| DTW | 0.741 | 0.76 | 0.734 |
| SVM | 0.860 | 0.842 | 0.851 |
| LSTM | 0.975 | 0.920 | 0.947 |
| **Bi-LSTM** | **0.956** | **0.940** | **0.948** |

**Table 2: Performance of different classification methods on the EgoGestAR dataset. Average of precision and recall values for all classes is computed to get a single number.**

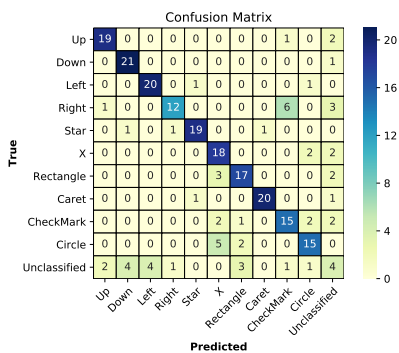

**Figure 4: The overall performance of *AirPen* on 240 egocentric videos captured using a smartphone device. The gesture is detected when the predicted probability is more than 0.85. Accuracy of our proposed framework is 0.8 (excluding the unclassified class).**

## Gesture Classification

We use the EgoGestAR, garg2018drawinair dataset for training and testing of the gesture classification network. LSTM and Bi-LSTM achieve classification accuracy of 92.5% and 94.3% respectively, outperforming the traditional approaches that are being used for similar classification tasks. We present comparison of the proposed LSTM and Bi-LSTM approach with DTW [9] and SVM [2] classification in Table 2. Additionally, we observe that the performance of traditional methods like DTW and SVM deteriorate significantly in the absence of sufficient data-points. Hence, they rely on complex interpolation techniques to give consistent results.

## Framework Evaluation

Since the proposed approach is a series of different networks, the overall classification accuracy in real-time will depend on the performance of each network used in the pipeline. Therefore, we evaluate the entire framework using 240 egocentric videos captured with a smartphone. The MobileNetV2 model was used in our experiments as it achieved the best trade-off between accuracy and performance.

The framework achieved an overall accuracy of 80.00% on a dataset of 240 egocentric videos as shown in Figure 4. The MobileNetV2 network works at 9 *FPS* on $640 \times 480$ resolution videos, and the *fingertip regressor* is capable of delivering frame rates of up-to 166 *FPS* working at a resolution of $99 \times 99$. The gesture classification network is capable of processing a given stream of data in less than 100*ms*. As a result, the average response time of the proposed framework is found to be 0.12*s* on a OnePlus 6 smartphone powered by a Snapdragon® 845 SoC. The entire model has a very small memory footprint of 16.3 *MB*.

We also perform a subjective evaluation for our framework. 40 subjects were chosen from our lab at random. Their ages ranged from $22 - 53$ years with diversity in culture and lifestyles. We prepared 5 use cases for our framework: as an interaction modality for a Google Cardboard based AR application, navigation while driving a car, watching a YouTube video while learning to cook something new, industrial inspection of 3D printer parts, and on a simulated assembly line. Each subject was assigned one task each and was asked to perform at least 3 instances of it. The score were averaged across instances and users. We notice that *AirPen* scores better than the traditional interaction modalities in all the use cases. There is significant disparity in scores for the Google Cardboard application. A possible explanation is that speech interfaces tend to be ineffective in ambient noise.

## DISCUSSION AND COMPARISON

Further analysis of the results shows that the *CheckMark* gesture is slightly correlated with the *Right* gesture since they both involve an arc that goes from left to right. Hence we observe a drop in the classification accuracy due to the inherent subjectivity of how a user draws a given gesture. We also make an important observation regarding our approach for including dedicated modules in our pipeline in contrast to an *end-to-end* framework.

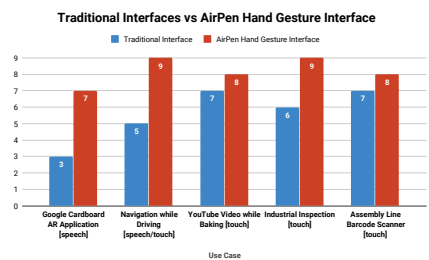

**Figure 5: Subjective evaluation of our proposed framework *AirPen*.** 40 subjects from our lab were assigned one task each and asked to rate the overall experience of using *AirPen* as opposed to the traditional interaction modality (mentioned in parenthesis) in 5 different use cases on a scale of 0 − 9. *AirPen* scores consistently better.

| Method | Accuracy | Time taken | On Device |
|---|---|---|---|
| Tsironi et al. | 32.27 | 0.76 | ✗ |
| VGG16+LSTM | 58.18 | 0.69 | ✗ |
| C3D | 66.36 | 1.19 | ✗ |
| **Proposed** | **80.00** | **0.12** | ✓ |

**Table 3: Analysis of gesture recognition accuracy and latency of various models against the proposed framework *AirPen*. Our framework works on-device and effectively has the highest accuracy and the least response time.**

We compared our modular pipeline against several *end-to-end* trained gesture classification methods Table 3. Tsironi et al.[19] proposed a network that works with differential image input to convolutional LSTMs to capture the body parts' motion involved in the gestures performed in second-person view. Even after fine-tuning the model on our egocentric video dataset, it produced an accuracy of only 32.14% as our data involved a dynamic background and no static reference to the camera. The VGG16+LSTM network [1] uses 2D CNNs to extract features from each frame. These frame-wise features are then encoded as a temporally deep video descriptor which is fed to an LSTM network for classification. Similarly, a 3D CNNs approach [18] uses 3D CNNs to extract features directly from video clips. Table 3 shows that both of these methods do not perform well. A plausible intuitive reason for this is that the network might be learning noisy and bad features while training. Our framework is currently limited to a single finger in the users' Field of View and the accuracy drops if multiple fingers are present at roughly the same distance. However, we observe that the framework performs equally well in cases where the user is wearing nail paint or has minor finger injuries. In such cases, the *fingertip regressor* outputs the point just below the nail and tracks it. Further, it is also robust to different hand colours and sizes. The EgoGestAR dataset can potentially be extended to a number of pointing gestures as per the requirement of the application since the framework can accommodate a number of sufficiently distinct gestures.

## CONCLUSION

We have presented *AirPen*: a framework to enable mass market reach of hand gestural interfaces that works in a resource constrained environment like on a smartphone or a video see-through HMD. Our approach is marker-less and uses only RGB data from a single smartphone camera. It works in real-time and on-device achieving an accuracy of 80.00% with a model size of 16.3 *MB*.

In the future, we intend to use our architecture to come up with applications of gesture recognition for people with a visual impairment or physical disability [20] and in situations of encumbrance [11]. Further, we aim to develop applications for huge display screens and boardrooms as an attempt to replace the mouse pointer with fingertip for controlling presentations.